\documentclass[APA,LATO1COL]{WileyNJDv5} 

\articletype{Article Type}%

\received{Date Month Year}
\revised{Date Month Year}
\accepted{Date Month Year}
\journal{Journal}
\volume{00}
\copyyear{2025}
\startpage{1}

\usepackage{algorithm, algpseudocode}

\newcommand{\bPhi}{\boldsymbol{\Phi}}
\newcommand{\bPsi}{\boldsymbol{\Psi}}
\newcommand{\bXi}{\boldsymbol{\Xi}}
\newcommand{\bb}{\boldsymbol{b}}
\newcommand{\ba}{\boldsymbol{a}}
\newcommand{\bc}{\boldsymbol{c}}
\newcommand{\bY}{\boldsymbol{Y}}
\newcommand{\bEps}{\boldsymbol{\epsilon}}
\newcommand{\bA}{\boldsymbol{A}}
\newcommand{\bC}{\boldsymbol{C}}
\newcommand{\bZ}{\boldsymbol{Z}}
\newcommand{\bD}{\boldsymbol{D}}
\newcommand{\bG}{\boldsymbol{G}}
\newcommand{\bL}{\boldsymbol{L}}
\newcommand{\bgamma}{\boldsymbol{\gamma}}
\newcommand{\cG}{\mathcal{G}}
\newcommand{\bI}{\boldsymbol{I}}       

\begin{document}

\title{Bayesian Functional Region Selection}

\author[1]{Hongxiao Zhu}

\author[1]{Yizhi Sun}

\author[1]{Jaeyoung Lee}

\authormark{ZHU \textsc{et al.}}
\titlemark{Bayesian Functional Region Selection}

\address[1]{\orgdiv{Department of Statistics}, \orgname{Virginia Tech}, \orgaddress{\state{Virginia}, \country{USA}}}

\corres{Hongxiao Zhu, Hutcheson Hall 406A, 250 Drillfield Dr, Blacksburg, VA 24061 \email{hongxiao@vt.edu}}



\abstract[Abstract]{Local regions on curves, images, and other high-dimensional objects often contain critical information for interpretation, prediction, and decision-making. Therefore, detecting local regions on functional data that are relevant to a variable of interest is highly desirable. We propose a Bayesian method for functional regression to select local regions on functional predictors that are relevant to a scalar response. The region selection is achieved through sparse estimation of the regression coefficient function. We adopt compactly supported and over-complete basis to capture local features of the coefficient function, and propose a spike-and-slab prior coupled with a structured Ising hyper-prior to encourage continuous shrinkage of nearby regions. Our proposed Bayesian framework accommodates both continuous and binary responses, resulting in posterior inference that naturally captures the uncertainty of the model parameters.}


\jnlcitation{\cname{%
\author{Zhu H.},
\author{Sun Y.},and
\author{Lee J.}}.
\ctitle{Bayesian functional region selection} \cjournal{\it Stat} \cvol{2024;00(00):1--18}.}

\maketitle



\section{Introduction}\label{sec:intro}

The ever-growing modern technology enables automatic collection of high-dimensional data in functional form, with basic observational units being curves, surfaces, or higher dimensional objects measured over fine grids. Typical examples include signals, images, and longitudinal trajectories. While functional data carry a large amount of information, in many applications the ``key'' information is usually conveyed by particular regions of functional data. For example, in medical imaging studies, it is often of interest to determine which lesions of the tissue on the PET or MRI images show abnormal changes. In pregnancy studies, researchers would like to know at which stage of the pregnancy process the progesterone level is influential in preterm birth. These questions can be generally summarized into one: how to identify regions of functional data that show a significant association with trait variables?

While it is possible to discretize functional observations into vectors and apply variable selection methods to vector features, the solution can be numerically unstable due to collinearity caused by the high correlation between nearby grid points in functional data. Additionally, there is no guarantee that meaningful regions will be identified by variable selection methods, as the spatial information of the grid points is often ignored.  It is thus highly desirable to develop new methods that encourage the selection of local regions while taking full advantage of the ``functional" nature of the data---treating functional data as realizations of random objects in a function space.

In the literature on functional data analysis, while considerable efforts have focused on functional regression \cite{Ramsay1997,Ferraty2006,Horvath2012,Morris2015,Wang2016}, only a few studies addressed the topic of region selection. In particular, James et al. \cite{james2009functional} used a $p$-dimensional basis to approximate the regression coefficient function $\beta(t)$ in a functional linear regression framework and applied a lasso-type penalty to the discretized approximation of the $d$th derivative, $\beta^{(d)}(t)$. Zhou et al. \cite{zhou2013functional} also studied functional linear regression and aimed to encourage zero values in the coefficient function at sub-regions. They proposed a two-stage estimator that combines a Dantzig selector with a group SCAD penalty. Koltchinskii and Minsker \cite{koltchinskii2013l_1} focused on $L_1$-norm penalization, a natural extension of Lasso, while assuming that the regression coefficient function is sparse. Lin et al. \cite{lin2015smooth} proposed a functional generalization of the SCAD penalty, named fSCAD, and combined it with smoothing splines to achieve smoothing and locally sparse estimation.

Despite their successes, these existing methods have several limitations. (i) Most existing methods rely on penalized regression within the Frequentist framework, and the uncertainty of the estimation is often hard to be quantified due to theoretical challenges. (ii) Existing methods typically use a single basis set to represent functional data, which is restrictive when the regions to be detected vary in scale. For example, most methods utilize B-splines with a fixed degree and a fixed number of knots to represent functional data. However, a fixed basis can be redundant for detecting regions at coarser scales while still insufficient for capturing finer-scale regions. (iii) Existing methods primarily focus on regressing a continuous scalar variable on a functional predictor and cannot be directly applied to other scenarios, such as those involving binary responses.

These limitations may be addressed by adopting a Bayesian approach. Several existing studies have used Bayesian models to identify local features in different contexts. For example, Zeng et al. \cite{zeng2024bayesian} proposed a image-on-scalar regression framework to model the global and local effects of scalar covariates on images. They utilize a pointwise spike-and-slab prior along with a hard-thresholding structure to achieve simultaneous ``global" covariate selection and ``local" pixel/voxel-level selection. In functional data analysis, 
several related studies on Bayesian functional mixed models have adopted multiple testing procedures based on discretized posterior samples of the regression coefficient functions to identify significant local regions \cite{Morris2008, Meyer2015, ZHU2018501, Huo2022}. These existing works demonstrate the potential of Bayesian approaches for region selection.  However, they are developed for specific scenarios such as image regression and functional-on-scalar regression. There remains a notable lack of Bayesian alternatives to the current Frequentist functional region selection methods.

In this paper, we address this research gap by introducing a Bayesian functional region selection model. The proposed model is based on the scalar-on-function regression framework and can identify regions relevant to a variable of interest. We employ compactly supported and potentially over-complete bases to capture local features of functional data across various scales, and then identify relevant regions through structured shrinkage of the regression coefficient function. Specifically, region selection is achieved through the continuous shrinkage of adjacent areas of the regression coefficient function utilizing a spike-and-slab prior coupled with an Ising hyper-prior. This yields a sparse estimate of the regression coefficient function, where regions with zero values correspond to irrelevant areas. A key advantage of the Ising hyper-prior is its ability to incorporate the neighboring structure of the compactly supported basis functions, which when combined with the spike-and-slab prior, ensures smooth and continuous shrinkage of the regression coefficient function. We propose Markov chain Monte Carlo (MCMC) algorithms for posterior inference, allowing us to fully capture the uncertainty of model parameters. Within our Bayesian framework, extending from continuous response to more complex cases, such as binary responses, becomes straightforward. The effectiveness of the proposed method is demonstrated through simulations and a real-world data application.

The remainder of this article is organized as follows. We provide details about the proposed Bayesian functional region selection method for the continuous response case in Sections \ref{subsec:flm}-\ref{subsec:flag} and for the binary response case in Section \ref{subsec:binary}. In Section \ref{sec:param}, we introduce approaches to determine hyper-prior parameters and improve computation efficiency. Simulation studies for both continuous and binary response cases are presented in Section \ref{sec:simulation}, followed by a real-world application using the Tecator data in Section \ref{sec:realdata}. Finally, a discussion is provided in Section \ref{sec:discuss}.

\section{Bayesian functional region selection}
\label{sec:region_selection}

\subsection{Functional linear model with compactly supported and over-complete basis}
\label{subsec:flm}

Let $Y_i$ denote a scalar response and $X_i(t)$ denote a functional predictor supported on a closed domain $T$ in $\mathbb{R}^d$, $d\geq 1$. Assume that $\{Y_i, X_i ;\; i=1,\dots, n\}$ are independent pairs following a {\it functional linear model} (FLM): 
\begin{eqnarray}
Y_i & = & b_0+ \int_{T} X_i(t)\beta(t)dt+\epsilon_i, \;\; \epsilon_i\sim N(0,\sigma^2),
\label{eq:flm}
\end{eqnarray}
where $X_i(t)$ and $\beta(t)$ are both functions in $L^2(T)$, the set of square integrable functions. Suppose that $\{X_i(t)\}$ are collected on a dense grid ${\bf t}=(t_1,\dots, t_{m})$ in $T$. We further assume that $\beta(t)$ is exactly zero on $T_0$, where $T_0 \subset T$, therefore only on the non-zero regions $T \setminus T_0$ is $Y_i$ associated with $X_i(t)$ via $\beta(t)$. At this stage, we assume that the functional predictors $\{X_i\}$ are observed without measurement errors; cases where $X_i$'s are subject to measurement errors are discussed in Section \ref{sec:discuss}.

We aim to achieve functional region selection by encouraging sparsity in the regression coefficient function $\beta(t)$, such that regions of $X_i(t)$ irrelevant to $Y_i$ correspond to the zero regions of $\beta(t)$. To identify sparse local regions of $\beta(t)$, it is necessary to capture the local structures of $\beta(t)$, which may be heterogeneous in smoothness, fluctuation frequency, etc., across different regions.  To this end, we represent $\beta(t)$ as a linear combination of a compactly supported basis in $L^2(T)$, denoted by $\boldsymbol{\Phi}=\{\phi_1,\dots,\phi_m\}$. Examples of such bases include Haar wavelets, Daubechies wavelets, B-splines, finite elements bases, etc. We allow the basis set $\boldsymbol{\Phi}$ to be potentially over-complete, meaning that it may contain more elements than necessary to span the functional space. The concept of over-complete bases stems from {\it sparse coding} or {\it dictionary learning} \cite{Olshausen1996}, where the goal is to find a sparse representation of data for more efficient compression or signal restoration. By using an over-complete basis, we expect local features of functional data to be more efficiently captured. For instance, when multi-scale bases are employed, features at various scales can be represented more parsimoniously with fewer basis functions compared to using a single set of bases at the finest scale.

\begin{figure*}[t]
\centerline{\includegraphics[height=9pc]{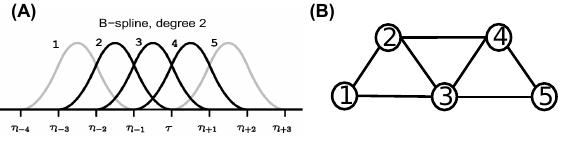}}
\caption{Visualization of the neighborhood structure of B-spline basis functions. (A) Five B-spline basis functions of degree two; (B) the corresponding neighborhood structure depicted as an undirected graph. \label{fig:Bspline}}
\end{figure*}

Depending on the nature of the functional data, different types of basis functions can be used in $\bPhi$. In this study, we use an over-complete B-spline basis as a demonstration example. In particular, we construct $\bPhi$ using multiple sets of B-spline bases at $L$ different levels of scale or degree. The scales of the B-spline bases can be controlled by the total number of knots on $T$, with fewer knots corresponding to a rougher scale. The degrees of B-splines are the degrees of polynomials used in constructing the B-spline bases. For example, a degree of zero corresponds to piecewise constant functions, while a degree of one corresponds to piecewise linear functions. We write $\bPhi= \left\{\phi_j^l(t), \, j=1,\ldots, q_l+k_l, \, l = 1,\ldots,L\right\}$, where $\phi_j^l$ denotes the $j$th B-spline basis function at level $l$, $q_l$ denotes the degree of the B-spline, and $k_l - 1$ equals the number of interior knots (in the one-dimensional setup). The total number of elements in $\bPhi$ is $p=\sum_{l=1}^L (q_l+k_l)$. Using this basis, we can express $\beta(t)$ as a linear combination of basis functions: $\beta(t) = \sum_{l=1}^L\sum_{j=1}^{q_l + k_l}b_j^l\phi_j^l(t)$, where $b_j^l$ is the coefficient corresponding to the $j$th basis component at level $l$. Plugging in the basis representation of $\beta(t)$ into model (\ref{eq:flm}), we get  
\begin{eqnarray}
Y_i = b_0 + \sum_{l=1}^L\sum_{j=1}^{q_l+k_l} \left(\int_{T} X_i(t)\phi_{j}^l(t)dt\right) b_j^l  + \epsilon_i.
\label{eq:expand}
\end{eqnarray}
If we denote $c_{ij}^l=\int_{T} X_i(t)\phi_{j}^l(t)dt$, then model (\ref{eq:expand}) can be reformulated as 
$Y_i=b_0 + \sum_{l=1}^L\sum_{j=1}^{q_l+k_l} c_{ij}^l b_j^l +\epsilon_i$.  We further concatenate $\{c_{ij}^l\}$ and $\{b_0, b_j^l\}$ into vectors, defining $\bb=(b_0, b_1^1,\ldots, b_{q_1 + k_1}^1, \ldots,b_1^L, \ldots, b_{q_L + k_L}^L )^T$ and $\bc_i=(1, c_{i,1}^1,\ldots,c_{i,q_{1} + k_{1}}^1,\ldots, c_{i,1}^L,\ldots,c_{i,q_{L} + k_{L}}^L)^T$.  Model (\ref{eq:expand}) then simplifies to $Y_i = \bc_i^T\bb + \epsilon_i$. Finally, we stack $\{\bc_i\}$ to form a matrix $\bC$, where $\bc_i^T$ is the $i$th row of $\bC$. Similarly, we stack $\{Y_i\}$ to form a vector $\bY = (Y_1, \ldots, Y_n)^T$ and stack $\{\epsilon_i\}$ to form a vector $\bEps = (\epsilon_1,\ldots,\epsilon_n)^T$. Using matrix-vector notation, model (\ref{eq:expand}) can be expressed as
\begin{eqnarray}
\bY = \bC \bb + \bEps.
\label{eq:compact_model}
\end{eqnarray}
While model (\ref{eq:compact_model}) appears similar to multiple linear regression, it differs fundamentally in two key ways. First, if $\bPhi$ is over-complete, the covariate matrix $\bC$ may become high-dimensional and exhibit multicollinearity among its columns. Thus, model (\ref{eq:compact_model}) constitutes a ``large p-small n"problem. Second, the basis functions in $\bPhi$ capture local information, which introduces a natural neighborhood structure among the columns of $\bC$. To illustrate this, Figure \ref{fig:Bspline} presents a simple example demonstrating the neighborhood structure induced by the basis functions. Figure \ref{fig:Bspline} (A) depicts five B-spline basis functions of degree two, assuming $L=1$. We define two basis functions as ``neighbors" if their supports overlap. This neighborhood structure can be represented by an undirected graph, as shown in Figure \ref{fig:Bspline} (B), where the nodes correspond to the basis functions and edges indicate neighborhood relationships. For an arbitrary point $t\in [\tau_{l-2},\tau_{l-1}]$, since $\beta(t)=\sum_{j=1}^5 \phi_j(t) b_{j}$, if $\beta(t)$ is exactly $0$ at $t$, the coefficients $\{b_j, j=1,2,3\}$ must all be zero simultaneously. In this study, we will encourage sparsity of $\beta(t)$ by imposing a prior on $\bb$ that leverages the underlying neighborhood structure, as described in detail in Section \ref{sec:prior}.

\subsection{Priors for structured variable selection}
\label{sec:prior}
As noted in Section \ref{subsec:flm}, we aim to encourage sparse estimation of $\beta(t)$ in model (\ref{eq:flm}), so that zero regions of $\beta(t)$ correspond to portions of $\{X_i(t)\}$ that are irrelevant to $Y_i$. This can be achieved by promoting a structured sparse estimation of $\bb$ in (\ref{eq:compact_model}) that takes into account the neighborhood relationships induced by $\bPhi$. To accomplish this, we propose a new prior for $\bb$
that combines the spike-and-slab prior \cite{Vannucci2012} with a structured Ising prior \cite{li2010bayesian}.
Specifically, denoting $b_j$ the $j+1$th component of $\bb$, we assume $b_0 \sim N(0, \sigma^2 v^2)$ and assume a spike-and-slab prior for $b_j$:
$b_j \mid \gamma_j, \sigma^2 \sim (1-\gamma_j)\delta_0 + \gamma_j N(0, \sigma^2 v^2)$, $j = 1,\ldots, p$,
where $\gamma_j \in \{0,1\}$, $\delta_0(\cdot)$ is a point mass at $0$. Here, $\bgamma = (\gamma_1,\ldots, \gamma_p)^T$ is a latent binary vector that determines whether the corresponding components in $\bb$ are zero or not. The parameter $\sigma^2 v^2$ is the prior variance of $b_j$ when $\gamma_j=1$, where $\sigma^2$ is the residual variance that serves as a scaling parameter. We usually set $v^2$ to be a large value (e.g., $1e+6$). In matrix form, the prior of $\{b_j\}$ can be expressed as  
$\bb \mid \bgamma, \sigma^2 \sim N\big(\boldsymbol{0}, \sigma^2\Sigma\big)$, where $\Sigma = \mbox{diag}\left\{v^2,\left(\gamma_j v^2 + (1-\gamma_j)\delta_0\right),j=1,\ldots,p\right\}$.

The spike-and-slab prior facilitates variable selection, resulting in sparse estimation of $\bb$. To further account for the neighborhood structure induced by $\bPhi$, we incorporate an Ising hyper-prior for the latent vector $\bgamma$, so that components in $\{\gamma_j\}$ corresponding to adjacent elements of $\bb$ are correlated a priori. In particular, let $\mathcal{G}=(\mathcal{V},\mathcal{E})$ denote the undirected graph that captures the neighborhood relationship induced by $\bPhi$, where $\mathcal{V}$ is the set of nodes and
$\mathcal{E}=\{(i,j): i\neq j, \, 1 \leq i,j\leq p\}$ is the set of edges. Additionally, let $\bG$ represent the (binary) $p\times p$ adjacency matrix corresponding to $\mathcal{G}$. An Ising hyper-prior for $\bgamma$ is written as
\begin{equation}
p(\boldsymbol{\gamma}) \propto \mbox{exp}\big\{\ba^T\boldsymbol{\gamma} + \boldsymbol{\gamma}^T \bD_{\cG}\boldsymbol{\gamma}\big\},
\label{eq:Ising}
\end{equation}
where $\boldsymbol{a} = \big(a_1,\ldots,a_p\big)^T$ is a real-valued vector and $\bD_{\cG} = \big(d_{i,j}\big)$ is a $p\times p$ symmetric matrix with zeros corresponding to those in $\bG$; that is, $d_{i,j} = 0$ if $(i,j)\notin \mathcal{E}$. In (\ref{eq:Ising}), $\ba$ is a hyperparameter that controls the sparsity of $\bgamma$ and $\bD_{\cG}$ is another hyperparameter that regulates the ``smoothness" of $\bgamma$. Specifically, the $(i,j)$th entry $d_{i,j}$ in $\bD_{\cG}$ represents the prior belief regarding the strength of coupling between the neighboring pair $(i,j)$, with larger values of $d_{i,j}$  indicating stronger coupling. We often assume that $d_{i,j}\geq 0$ which corresponds to the situation that adjacent elements of $\bgamma$ tend to have the same sign. In the extreme case, when $\bD_{\cG}\equiv \boldsymbol{0}$, the Ising prior reduces to the product of independent Bernoulli priors for elements of $\bgamma$. Finally, if no prior information is available for setting different components of $\ba$ and $\bD_{\cG}$, a simplified setup can be adopted by letting $\ba = a \, (1,\dots, 1)^T$ and $\bD_{\cG} = d \,\bG$. This reduces the parameters in (\ref{eq:Ising}) from $(\ba, \bD_{\cG})$ to two scalars $(a, d)$. The resulting hyper-prior density is
$p(\boldsymbol{\gamma}) \propto \mbox{exp}\big\{a\mathbf{1}^T\boldsymbol{\gamma} + d\,\boldsymbol{\gamma}^T \bG \boldsymbol{\gamma}\big\}$, $a\in \mathbb{R}, d\geq 0$.

In addition to the prior for $\bb$, we adopt an inverse-Gamma prior for the residual variance $\sigma^2$, specifically $\sigma^2 \sim \mbox{Inv-Gamma}(u/2,u\lambda/2)$. In the special case when $u = 0$, the inverse-Gamma prior reduces to a diffuse prior with density $p(\sigma^2) \propto 1/\sigma^2$. These prior parameterizations enable the sequential integration of $\boldsymbol{b}$ and $\sigma^2$ from the joint posterior distribution, thereby enhancing the efficiency of posterior inference.

\subsection{Posterior inference}
\label{sec:post}
Based on the model representation in (\ref{eq:compact_model}) and the prior setups in Section (\ref{sec:prior}), we can derive the marginal posterior distribution $p(\bgamma|\bY)$ as well as the conditional posterior distributions $p(\sigma^2|\bgamma,\bY)$ and $p(\bb|\sigma^2,\bgamma,\bY)$.  Given $\bgamma$, let $\bb_{\bgamma}$ denote the sub-vector of $\bb$ corresponding to the non-zero components of $\bgamma$ along with the intercept term. With this notation, the prior for $\bb_{\bgamma}$ can be expressed as $N(\boldsymbol{0}, \sigma^2 v^2 \bI_{|\bgamma|+1})$, where $\bI_{|\bgamma|+1}$ is an identity matrix of size $|\bgamma|+1$ and $|\bgamma|$ represents the number of ones in $\bgamma$. Similarly, let $\bC_{\bgamma}$ denote the submatrix of $\bC$ whose columns correspond to the non-zero entries of $\bb$, along with an additional column of ones for the intercept. The likelihood of $\bY$ given $\bgamma, \bb_{\bgamma}$ and $\sigma^2$ is
$\bY \mid \bb_{\bgamma}, \bgamma, \sigma^2 \sim N\big(\bC_{\bgamma}\bb_{\bgamma}, \sigma^2 \bI_n\big)$. 
The joint posterior distribution of $\bb_{\bgamma}$, $\bgamma$ and $\sigma^2$ can be written as
$p(\bb_{\bgamma},\bgamma,\sigma^2 \mid \bY) \propto p(\bY \mid \bb_{\bgamma}, \bgamma, \sigma^2)\,
p(\bb_{\gamma} \mid \bgamma,\sigma^2)\, p(\bgamma)\, p(\sigma^2)$.
Assuming that $p(\sigma^2)\propto 1/\sigma^2$, we first derive the marginal likelihood $p(\bY\mid \bgamma)$ by integrating out $\bb_{\bgamma}$ and $\sigma^2$ sequentially from the joint distribution $p(\bY, \bb_{\bgamma}, \sigma^2 \mid \bgamma) \propto p(\bY\mid \bb_{\bgamma}, \bgamma, \sigma^2)\,p(\bb_{\gamma} \mid \bgamma,\sigma^2)\, p(\sigma^2)$. This results in
\begin{eqnarray}
p\big(\bY \mid \bgamma \big) =\frac{\Gamma\big(\frac{n}{2}\big)}{(\pi)^{\frac{n}{2}}} \frac{ \left|v^2 \bC_{\bgamma}^T \bC_{\bgamma}+\bI_{|\bgamma|+1}\right|^{-\frac{1}{2}}}
 {\big(\bY^T \bL_{\bgamma}\bY\big)^{\frac{n}{2}}}.
\label{eq:marginal_post}
\end{eqnarray}
Here, $\bL_{\bgamma} = \bI_{n} -\bC_{\bgamma}\big(\bC_{\bgamma}^T\bC_{\bgamma}+v^{-2}\bI_{|\bgamma|+1}\big)^{-1}\bC_{\bgamma}^T$, and $\bI_{|\bgamma|+1}$ and $\bI_{n}$ are identity matrices of size $|\bgamma|+1$ and $n$, respectively. As a by-product of the integration, we obtain the conditional posterior distribution of $\bb_{\bgamma}$ as 
\begin{eqnarray}
\bb_{\bgamma} \mid \sigma^2, \bgamma, \bY \sim N\big(\mu_{\bb}, V_{\bb}\big),
\label{eq:bgamma}
\end{eqnarray}
 where $\mu_{\bb} = \big(\bC_{\bgamma}^T\bC_{\bgamma} + v^{-2}\bI_{|\bgamma|+1}\big)^{-1}\bC_{\bgamma}^T\bY$ and $V_{\bb} = \sigma^2 \big(\bC_{\bgamma}^T\bC_{\bgamma} + v^{-2}\bI_{|\bgamma|+1}\big)^{-1}$. It is noteworthy that $p(b_j|\sigma^2, \gamma_j=0, \bY)=\delta_0$, representing a point mass at $0$ for $j=1,\dots, p$. Additionally, the conditional posterior distribution of $\sigma^2$ is given by
\begin{eqnarray}
\sigma^2 \mid \bgamma,\bY \sim \mbox{Inv-Gamma}(\widetilde{\alpha}, \widetilde{\beta}),
\label{eq:sigma2}
\end{eqnarray}
where $\widetilde{\alpha} = n/2$ and $\widetilde{\beta} = \bY^T \bL_{\bgamma}\bY/2$.

These posterior distribution results enable us to design a Markov chain Monte Carlo (MCMC) algorithm using the Gibbs sampler. Specifically, we first update each component of $\bgamma$ sequentially based on the conditional posterior odds, which takes the form 
\begin{equation}
\mbox{Odds}_j = \frac{p\big(\gamma_j = 1 \mid \bgamma_{(-j)}, \bY\big)}{p\big(\gamma_j = 0 \mid \bgamma_{(-j)}, \bY\big)}= \frac{p\big(\bY\mid\gamma_j = 1,\bgamma_{(-j)}\big)}{p\big(\bY\mid\gamma_j = 0,\bgamma_{(-j)}\big)}
\cdot \frac{p\big(\gamma_j = 1 \mid \bgamma_{(-j)}\big)}{p\big(\gamma_j = 0 \mid \bgamma_{(-j)}\big)}, \quad j=1,\dots,p,
\label{eq:odds}
\end{equation}
where $\bgamma_{(-j)} = \big\{\gamma_{l},\, l \neq j\big\}$. Each $\gamma_j$ can be updated by sampling from a Bernoulli distribution with probability $\mbox{Odds}_j/(\mbox{1+Odds}_j)$. Given the updated $\bgamma$, we further update $\sigma^2$ and $\bb_{\bgamma}$ from their conditional distributions.
Note that the prior odds in (\ref{eq:odds}) takes the form  
\begin{equation}
\frac{p\big(\gamma_j = 1 \mid \boldsymbol{\gamma}_{(-j)}\big)}{p\big(\gamma_j = 0 \mid \boldsymbol{\gamma}_{(-j)}\big)}
= \frac{p\big(\gamma_j = 1,\boldsymbol{\gamma}_{(-j)}\big)}{p\big(\gamma_j = 0,\boldsymbol{\gamma}_{(-j)}\big)}
= \mbox{exp}\big\{a_j + 2 \boldsymbol{d}_{j}^T \bgamma \big\}, \quad j = 1,\ldots,p,
\label{eq:prior_odds}
\end{equation}
where $\boldsymbol{d}_{j}$ is the $j$th column of $\bD_{\cG}$. If we adopt the simplified two-parameter formulation of Ising prior, the prior odds simplifies to $\mbox{exp}\big\{a + 2 d \, \boldsymbol{g}_j^T \bgamma \big\}$ for $j = 1,\ldots,p$, where $\boldsymbol{g}_j$ is the $j$th column of the adjacency matrix $\bG$. We outline the steps of the MCMC algorithm for region selection in functional linear models in Algorithm \ref{ag:flm}. 
\begin{algorithm}
\caption{\enskip An MCMC algorithm for region selection in functional linear models}\label{ag:flm}
\begin{algorithmic}[1]
\State Initialize the parameter $\bgamma$. 
\For{ $j=1,\dots,p$ }
        \State Conditional on $\bgamma_{(-j)}$, update $\gamma_j\sim \mbox{Bernoulli}(\frac{\mbox{Odds}_j}{1+\mbox{Odds}_j})$, where $\mbox{Odds}_j$ is calculated based on (\ref{eq:odds}). 
        \EndFor 
\State Conditional on $\bgamma$, update $\sigma^2$ based on (\ref{eq:sigma2}).
\State Conditional on $\bgamma$ and $\sigma^2$, update $\bb_{\bgamma} $ based on (\ref{eq:bgamma}) and $(b_j|\sigma^2, \gamma_j=0, \bY)=0$. 
\State Repeat steps 2-6 until the maximum number of iterations is reached. 
\end{algorithmic}
\end{algorithm}

\subsection{Flag regions based on posterior samples}
\label{subsec:flag}
Based on posterior samples of $\bb$, we can reconstruct posterior samples of $\beta(t)$ via the expansion $\beta(t) = \sum_{l=1}^L\sum_{j=1}^{q_l + k_l}b_j^l\phi_j^l(t)$. While zero regions can be identified by naively flagging positions on which the posterior mean of $\beta(t)$ is zero, we will flag regions by taking into account posterior variability following the Bayesian multiple testing procedure proposed by Morris et al. and others \cite{Morris2008, Meyer2015, ZHU2018501, Huo2022}. For the testing procedure, since $\beta(t)$ is continuous on $T$, it is reasonable to treat $\beta(t)$ as zero when $|\beta(t)| \leq \delta$ for a small $\delta$. Therefore, instead of testing whether $|\beta(t)|>0$, we will test whether $|\beta(t)|>\delta$ for a pre-specified value of $\delta$. For example, one could set $\delta=1e-3$. One can also choose larger values for $\delta$ if desired. To perform the testing, we consider a sequence of tests indexed by $t$ on a dense grid {\bf t}=$(t_1, \ldots, t_m)$ of $T$, with null hypotheses $\{H_0: |\beta(t)| \leq \delta, t \in {\bf t}\}$. The Bayesian multiple testing procedure involves three steps: first, we calculate the point-wise posterior probability $\widehat{p}(t) \approx \mbox{pr}(|\beta(t)| > \delta\mid\mbox{Data})$ from the posterior samples of $\beta(t)$; second, we find a threshold $\phi_{\alpha}$ for $\widehat{p}(t)$ by controlling the Bayesian expected false discovery rate (FDR) to be less than $\alpha$; finally, we flag locations where $\widehat{p}(t) > \phi_{\alpha}$ as being significantly greater than $\delta$. We refer to the above region flagging approach as the Bayesian false discovery rate (BFDR) approach. Further details of this testing approach are provided in the online supplementary materials. 

\subsection{Extension to the binary response case}
\label{subsec:binary}
If the response variable $Y_i$ is binary, we introduce a latent variable $Z_i$ that connects the response to the functional predictor via a probit link. Specifically,  
\begin{eqnarray}
  Y_i &= & \left \{
  \begin{aligned}
    &1, && \text{if}\ Z_i < 0  \nonumber\\
    &0, && \text{if}\ Z_i \geq 0 
  \end{aligned} \right. \\
 Z_i & = & b_0+ \int_{T} X_i(t)\beta(t)dt + \epsilon_i, \quad \epsilon_i \sim N(0,1),   \label{eq:latent}
\end{eqnarray}
for $i = 1,\ldots,n$. Note that due to the use of the probit link, we no longer need $\sigma^2$ in the variance of the residual error $\epsilon_i$. We follow the same idea as in Section \ref{subsec:flm} to represent $\beta(t)$ using a compactly supported basis, which transforms the linear part of the above model to $Z_i = \bc_i^{T}\bb + \epsilon_i$.  Under this setup, we assume the same spike-and-slab prior for $\bb$ and an Ising hyper-prior for the binary latent vector $\bgamma$.

This framework allows us to derive the conditional distribution of the latent variable $Z_i$ given $Y_i$ and $\bb$ as a truncated normal (TN) distribution:
\begin{equation}
Z_i \mid Y_i, \mathbf{b} \sim \mbox{TN}\big(\bc_i^T\bb, 1\big)\left\{I_{\{Z_i < 0\}}I_{\{Y_i = 1\}} + 
I_{\{Z_i \geq 0\}}I_{\{Y_i = 0\}}\right\}. 
\label{eq:tn}
\end{equation}
Denote $\bZ=(Z_1,\ldots, Z_n)^T$. The conditional distribution of $\bZ$ given $\bgamma$ and $\bC_{\bgamma}$ can be derived by integrating out $\bb$, i.e., 
\begin{equation}
p(\bZ \mid \bgamma, \bC_{\bgamma}) \propto v^{-(|\bgamma|+1)}|\bA|^{-1/2}\exp\{-\bZ^T\bL_{\bgamma}\bZ/2\},
\label{eq:cond_z}
\end{equation}
where $\bA = \bC_{\bgamma}^T\bC_{\bgamma} + v^{-2}\bI_{|\bgamma|+1}$ and $\bL_{\bgamma} = \bI_{n} -\bC_{\bgamma}\bA^{-1}\bC_{\bgamma}^T$. A Gibbs sampler that iteratively samples from $p(\bZ \mid \bY, \bb)$, $p(\bgamma \mid \bZ, \bC_{\bgamma})$, and $p(\bb \mid \bgamma, \bC_{\bgamma}, \bZ)$ can be designed. We describe the steps of the Gibbs sampler in Algorithm \ref{ag:probit}. 

\begin{algorithm}
\caption{\enskip A Gibbs sampler for region selection in binary-response functional-predictor models}\label{ag:probit}
\begin{algorithmic}[1]
\State Initialize the parameters $\bgamma$, $\bb$; 
\State Given $\bb$, update $\bZ$ based on (\ref{eq:tn}); 
\State Given $\bZ$, update $\bgamma$ by calculating posterior odds based on (\ref{eq:cond_z}) and (\ref{eq:prior_odds}); 
\State Given $\bZ$ and $\bgamma$, update $\bb\sim N(\mu_{\bb},V_{\bb})$, where $\mu_{\bb}=\bA^{-1}\bC_{\bgamma}^T\bZ$, and $V_{\bb}=\bA^{-1}$;      
\State Repeat steps 2-4 until the maximum number of iterations is reached. 
\end{algorithmic}
\end{algorithm}

\section{Model Implementation}
\label{sec:param}
To implement the proposed model, we need to specify the set of basis functions and their neighborhood structure. We also need to set prior parameters and choose initial values for the algorithms. Additionally, since the algorithm involves selecting variables in a high-dimensional space, techniques to improve computational efficiency are desirable. We will discuss these aspects in this section.

\subsection{Specify basis functions and their neighborhood structure}
\label{subsec:basis}
The basis set $\bPhi$ should satisfy two conditions: (i) it should fully span the function space $L^2(T)$, and (ii) it should capture local regions at the desired scale. ``Global" bases that are supported on the entire domain $T$, such as functional principal component and Fourier bases, are non-zero almost everywhere. As a result, they cannot capture local features effectively and therefore do not meet condition (ii). A single set of B-spline bases with a fixed degree and number of knots offers local support but may lack the resolution needed for  regions with finer-scale features while becoming redundant in areas where only coarser-scale features are present. 
We therefore recommend using a collection of B-spline bases with varying numbers of knots and/or degrees, or alternatively, a single basis with multiresolution properties, such as Daubechies wavelets. Given the basis set, the neighborhood structure (i.e., the adjacency matrix $\boldsymbol{G}$) can be determined as follows. First, for each pair of basis functions $\{\phi_l (t), \phi_{l'}(t)\}$, we identify their supports by flagging out grid points on which the absolute function value exceeds a small threshold $\delta_1$ (e.g., $\delta_1 = 1e-2$); we denote these support sets as $\boldsymbol{t}_l$ and $\boldsymbol{t}_{l'}$, respectively. Two basis functions are considered to overlap if $\boldsymbol{t}_l\cap\boldsymbol{t}_{l'}\neq \varnothing$. If overlap is present, we set $G_{l,l'} = 1$ if the ratio $|\boldsymbol{t}_l\cap\boldsymbol{t}_{l'}|/\mbox{min}\{|\boldsymbol{t}_l|,|\boldsymbol{t}_{l'}|\}$ exceeds a prespecified percentage $\delta_2$ (e.g., $\delta_2=30\%$),  where $|\cdot|$ denotes set cardinality. Larger $\delta_2$ results in a sparser $G$. We can obtain a reasonable neighborhood relationship between the basis functions by adjusting $(\delta_1, \delta_2)$. 

Our proposed method requires initial values for $\bgamma$ in Algorithm \ref{ag:flm} and for both $\bgamma$ and $\bb$ in Algorithm \ref{ag:probit}. The initial values of $\bgamma$ can be set in two ways. One straightforward approach is to randomly sample each component of $\bgamma$ independently from a Bernoulli distribution, with the parameter determined by a prespecified sparsity level of $\bgamma$; for example, $\gamma_i\sim \mbox{Bernoulli}(0.1)$ for $i=1,\dots, p$. Alternatively, a more sophisticated approach involves
determining the initial value of $\bgamma$ based on a preliminary estimate of $\beta(t)$. This preliminary estimate, denoted by $\beta^0(t)$, can be obtained by fitting a classical functional regression model that does not assume zero regions, such as functional linear models \cite{ramsay2009introduction, yao2005functional} or generalized functional linear models \cite{mueller2005}. Using ${\beta}^0(t)$, one can obtain an initial value for $\bb$ by regressing ${\beta}^0(t)$ on the basis functions $\{\phi_j^l(t)\}$. For example, one can fit ${\beta}^0(t_i) = \sum_l\sum_j\phi_j^l(t_i) b_j^l +\epsilon_i, i=1, \dots, m$, while applying 
a lasso penalty on $\{b_j^l\}$. 
This results in an initial estimate for $\bb$. The initial value of $\bgamma$ can be obtained by applying a hard threshold to $\bb$. Our observation indicates that, while good initial values often facilitate faster convergence, posterior results are generally not sensitive to the choice of initial values, provided that a sufficient number of MCMC iterations are performed. 


\subsection{Set parameters for the Ising hyper-prior}
\label{subsec:ising}
The choice of Ising hyper-prior parameters $(\ba,\bD_{\mathcal{G}})$ directly controls the sparsity of $\bgamma$ as well as the strength of coupling between elements in $\bgamma$, which further influences the model size and computation time. When specifying these parameters, we also need to avoid the potential {\it phase transition} problem, which will be discussed in more detail in this section. As we often do not have a priori information regarding the different components of $\bgamma$, specifying each element of $\ba$ and $\bD_{\mathcal{G}}$ individually is challenging. Therefore, we focus on the simplified setup $\ba = a \, (1,\dots, 1)^T$ and $\bD_{\cG} = d \,\bG$, which only requires specifying two scalar prior parameters $(a,d)$.

We introduce several empirical approaches to help us determine the Ising parameters. The first two steps listed below provide a rough range of $a$ and $d$, while the third step offers empirical samples to assist us in selecting from a list of potential $(a,d)$ combinations and in avoiding the phase transition issue. 

\begin{enumerate}
\item {\it Determine an upper bound for $a$.} We start with an initial estimate of the sparsity level of $\bgamma$, denoted by $\widehat{q}_1$, which reflects our prior belief about the proportion of ones in $\bgamma$. If we have an initial estimate $\widehat{\bgamma}$, $\widehat{q}_1$ can be estimated by $\sum_{j=1}^p \widehat{\gamma}_j/p$. To find an upper bound for $a$, we set $d=0$; thus, the Ising prior reduces to independent Bernoulli distributions with odds $\exp\{a\}$ for each $\gamma_i$. We simply set $\widehat{q}_1=1/(1+\exp\{-a\})$ and solve for $a$, yielding $a=\mbox{logit}(\widehat{q}_1)$. Note that if we expect the marginal probability of $\gamma_i=1$ to be less than $0.5$, then $a$ should be a negative number, which is often the case when an over-complete basis is used. We note that $\mbox{logit}(\widehat{q}_1)$ is indeed an upper bound for $a$ because, as $d$ increases from $0$, it also increases the probability that each $\gamma_i$ equals $1$, a phenomenon referred to as the ``attractive pulling" effect \cite[p.~14]{Weller2014}. As a result, as $d$ increases, we will need a smaller $a$ to achieve the same marginal sparsity level. Therefore, we shall set $a\leq\mbox{logit}(\widehat{q}_1)$. 
 
\item {\it Determine the range of $d$ given $a$.} Since the parameter $d$ controls the strength of the attractive pull between neighboring elements in $\bgamma$, we can determine $d$ by controlling $\mbox{pr}(\gamma_i=1|\gamma_j=1, j\in \mathcal{N}(i))$, which characterizes the probability that $\gamma_i$ equals one given that all its neighbors are ones. Here, $\mathcal{N}(i)$ denotes the set of neighbors of node $i$. We observe that 
$\mbox{pr}(\gamma_i=1|\gamma_j=1, j \in \mathcal{N}(i))=(1+\exp\{-a - 2 d\sum_{j=1}^p G_{ij}\})^{-1}$. One can set this probability to be $\widehat{q}_2$ to control the strength of the attractive pull between neighboring elements. A higher $\widehat{q}_2$ indicates a stronger pulling effect, resulting in smoother $\bgamma$. We note that $\widehat{q}_2$ should not be set too close to one as suggested by McEvoy et al. \cite{McEvoy2013}. Solving for $d$ gives $d=(\mbox{logit}(\widehat{q}_2)-a)/(2\sum_{j=1}^p G_{ij})$. Note that the denominator depends on $i$. We can choose the mode across indices $i$ to get $d=(\mbox{logit}(\widehat{q}_2)-a)/(2 \, \mbox{mode} \, (\sum_{j=1}^p G_{ij}))$, which accounts for the most common neighborhood structure. In an extreme case, given a fixed $a$, if we set $\widehat{q}_2=\widehat{q}_1$, we are assuming conditional independence, which corresponds to the minimum pulling strength, i.e., $\mbox{pr}(\gamma_i=1|\bgamma_j=1, j\in \mathcal{N}(i))=\mbox{pr}(\gamma_i=1)$. This should result in a lower bound for $d$. 

\item {\it Run MCMC to determine $(a,d)$.} Steps 1-2 provide boundaries or ranges for $(a,d)$. To set specific values for $(a,d)$, we need to simultaneously control properties such as the sparsity level and the strength of the pulling. This can be achieved by performing MCMC Gibbs sampling based on the conditional distribution $\mbox{pr}(\gamma_i=1|\gamma_j=1, j\neq i)$. The samples of $\bgamma$ provide empirical estimates of the sample mean and sample correlation of the Ising distribution. 
These statistics provide valuable guidance for choosing appropriate values of $(a,d)$.
\end{enumerate}

\begin{figure} 
\begin{center}
\includegraphics[width=4.0 in]{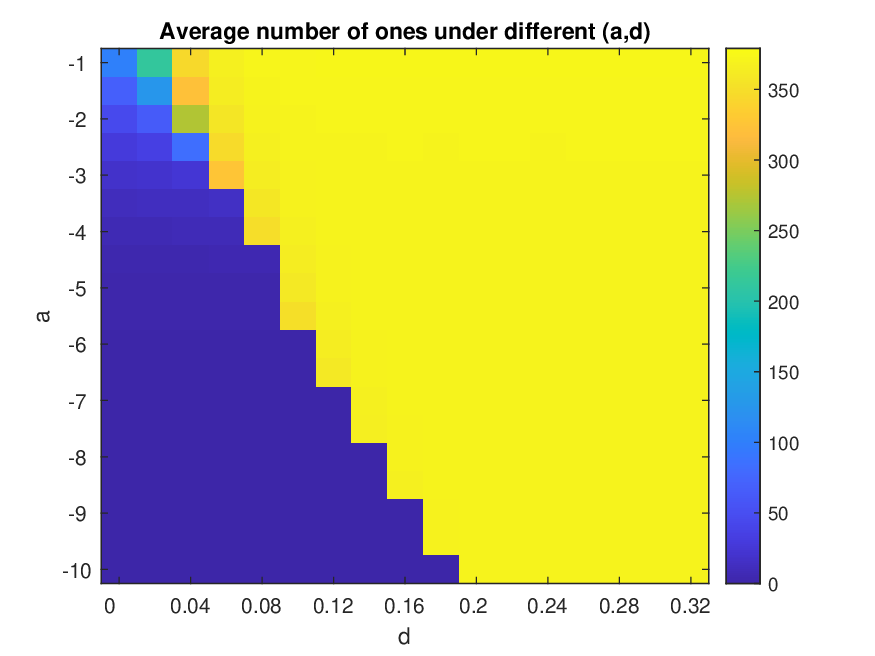}
\caption{A heatmap for the statistic $\sum_i \gamma_i$ under different $(a,d)$ combinations, obtained by sample averages.}
\label{heatmap}
\end{center}
\end{figure}

In addition to empirical moments, the prior samples of $\bgamma$ also provide the empirical distribution for the statistic $\sum_i \gamma_i$---the number of ones in $\bgamma$. This statistic can help us avoid the phase transition issue. Phase transition is a phenomenon that occurs when a small change in the parameter leads to a large-scale, qualitative change in the state of a system. If a small change in $(a,d)$ results in a rapid increase of $\sum_i \gamma_i$ from nearly zero to nearly $p$, it indicates that $(a,d)$ is on the boundary of a phase transition. Phase transition can lead to several serious consequences, such as a dramatic change in the proportion of $\gamma_i= 1$ when performing Bayesian variable selection for high-dimensional $\bgamma$, a slowdown in computation, or non-convergence of the MCMC algorithm. Therefore, we should select values of $(a,d)$ to avoid regions of phase transition. In Figure~\ref{heatmap}, we present a heatmap of the sample mean of the statistic $\sum_i \gamma_i$ at different $(a,d)$ combinations. In this demonstration, the graph $\boldsymbol{G}$ is calculated based on $379$ B-spline basis functions used in Simulation 1. From Figure~\ref{heatmap}, we can clearly identify $(a,d)$ values that are close to the boundary between the blue and yellow regions, which corresponds to the transition from sparse to dense $\bgamma$. To check whether phase transition occurs in the Ising prior, we can examine the histogram of the statistic $\sum_i \gamma_i$ to determine whether it has two modes: one near zero and one near $p$. 

\subsection{Improve computational efficiency}
\label{subsec:comput}
While an over-complete basis offers substantial advantages in region selection, it also introduces computational costs due to the need to perform stochastic search in a much higher-dimensional space. Bayesian variable selection in high dimensions has been extensively studied~\cite{OHara2009,Vannucci2012,LuLou2022}. Below, we outline several techniques that we have found useful for improving the computational speed and mixing of the MCMC chain in the context of functional region selection. 

\begin{itemize}
\item Randomize the order of updating elements of $\bgamma$ in Gibbs sampling. When updating $\bgamma$ in Algorithms \ref{ag:flm} and \ref{ag:probit}, we randomly permute the indices of the elements in $\bgamma$ and update $(\gamma_i | \gamma_j, j\neq i)$ in this randomized order. Studies have shown that randomization can significantly affect the mixing time for certain models~\cite{HeNIPS2016}.  
\item Carry over the log-likelihood under the current $\bgamma$ during MCMC iterations. Using the functional linear model as an example, when updating $\bgamma$, we need to calculate the Bayes factor $f(\bY|\gamma_j=1, \bgamma_{(-j)})/f(\bY|\gamma_j=0, \bgamma_{(-j)})$ in log scale for each $j$. To avoid recalculating both the numerator and denominator during every iteration, we can save the log-likelihood for the updated $\gamma_j$ at each step and carry it over to the next iteration. This approach can reduce computation time by nearly half. The same technique can also be applied when updating $\bgamma$ in the binary-response model.
\item Use the sweep operator. In the proposed algorithms, the most computationally expensive step is calculating the conditional log-likelihood based on Equation (\ref{eq:marginal_post}). We can rewrite the density as 
\begin{eqnarray}
p\big(\bY \mid \bgamma \big) \propto v^{-(|\bgamma|+1)} \frac{ \left|\bA\right|^{-\frac{1}{2}}}{\big(\bY^T\bY-\bY^T \bC_{\bgamma}\bA^{-1}\bC_{\bgamma}^T\bY\big)^{\frac{n}{2}}},
\label{eq:rewrite}
\end{eqnarray}
where $\bA=\bC_{\bgamma}^T \bC_{\bgamma}+v^{-2}\bI_{|\bgamma|+1}$. By using the sweep operator, we can simultaneously compute the log of the numerator and denominator in Equation (\ref{eq:rewrite}) with
a single sweep of the following matrix across the first block:  
\[\begin{bmatrix}
\bA & \bC_{\bgamma}^T\bY \\
\bY^T \bC_{\bgamma} & \bY^T\bY 
\end{bmatrix}.\]
More details about the sweep operator can be found in Wolfinger et al.~\cite{Wolfinger1994}.   
\end{itemize}
In addition to the computational techniques mentioned above, we also implemented a simple prescreening step to exclude columns of $\bC$ that are zero or close to zero, as such columns do not contribute to the regression due to their low magnitude. This step helps reduce the dimensionality of the search space, thereby alleviating the computational burden.

\section{Simulation studies}
\label{sec:simulation}

We conducted two simulation studies to evaluate the performance of our proposed approach: one for the continuous response case and the other for the binary response case. Both studies utilized the same set of functional predictors, $\{X_i(t), i=1,\dots, n\}$, with $n = 300$. The functional predictors were smooth trajectories sampled from a Gaussian process with a mean function $\mu(t) = \sqrt{16t} + 0.6\sin(16t) - 1.5$ and a correlation kernel $\kappa(s,t)=2^{1-\nu}(2|s-t|\sqrt{\nu}/\alpha)^{\nu}K_{\nu}(2|s-t|\sqrt{\nu}/\alpha)/\Gamma(\nu)$ in the Mat{\'e}rn family~\cite{Handcock94}. Here, $K_{\nu}(\cdot)$ represents the modified Bessel function of order $\nu$. We set $\nu=0.5$, $\alpha=1.5$ in the kernel and set the marginal variance at each grid point to be $1.2$. The predictor trajectories were sampled on a dense, equally spaced grid over the interval $T = [0,1]$, consisting of $m=400$ grid points. In addition to the functional predictors, the true regression coefficient function was defined as $\beta(t) = \cos(4\pi t)\, I_{[0,\,0.375]}(t) + \cos(16\pi t)\, I_{[0.719,\, 0.781]}(t)$. A local smoothing was further applied to $\beta(t)$ using moving average with a bin width of $20$ grid points. The resulting coefficient function $\beta(t)$ features two distinct non-zero regions, located at $[0, 0.375]$ and $[0.719, 0.781]$, as illustrated by the red dashed lines in Figure~\ref{simu_cont_response}. In the binary response case, the functional predictors $\{X_i(t)\}$ from the continuous response case were scaled by a factor of $30$ to enhance the signal-to-noise ratio. Binary responses were then generated by first sampling $Z_i \sim N\big(\int_{T} X_i(t)\beta(t)dt, 1\big)$ and subsequently defining $Y_i = I_{\{Z_i < 0\}}$.

Our goal is to compare the performance of the proposed method with several alternative models in terms of both estimation accuracy and the ability to detect non-zero regions. In the continuous response case, we considered four alternative models---two that do not assume zero regions on $\beta(t)$ and two that do. Specifically, we compared with the Functional Linear Model with B-splines (FLM-Bspline) \cite{ramsay2006functional}, the Functional Linear Model through Principal Analysis by Conditional Expectation (FLM-PACE) \cite{yao2005functional}, the Functional Linear Regression That's Interpretable (FLiRTI) \cite{james2009functional}, and the Smooth and Locally Sparse (SLoS) model \cite{lin2015smooth}. 
In the binary response case, since no existing approaches encourage zero regions of $\beta(t)$, we compared the proposed method with two versions of Generalized Functional Linear Models (GFLMs)---one using the basis representation approach with B-splines (GFLM-Bspline) \cite{Febrero2012} and the other using Functional Principal Component Analysis based on Principal Analysis by Conditional Expectation (GFLM-PACE) \cite{mueller2005}. Note that, since all the comparison methods are Frequentist methods, uncertainty characterization is typically unavailable. To characterize uncertainty, we performed $1000$ bootstraps on the original simulated data and used the estimates of $\beta(t)$ based on bootstrapped samples to create confidence bands. Using these estimates of $\beta(t)$, we also conducted BFDR testing as described in Section \ref{sec:post} to detect non-zero regions in $\beta(t)$. This approach enables us to compute the same summary statistics for both estimation and region detection across all methods, facilitating a direct side-by-side comparison. In addition to estimating $\beta(t)$, we also assessed the prediction performance of each method on an independently simulated test set. The results indicate that all methods produce comparable prediction results. 
Therefore, we do not report the prediction performance in the following subsections.


\subsection{The continuous response case}
In the continuous response case, $Y_i$ is generated from $N\big(\int_{T} X_i(t)\beta(t)dt,\, \sigma^2\big)$ with $\sigma=0.05$. Here, the integral was evaluated using numerical integration via the trapezoidal rule. To evaluate the performance of the proposed region selection method under different basis configurations, we employed two settings of multiresolution B-spline bases $\bPhi$, one resulting in a lower dimensional setting and the other resulting in a higher-dimensional setting. In the lower dimensional setting, we used B-splines of degrees two and three ($q_1$=2, $q_2=3$), corresponding to quadratic and cubic polynomials, respectively. Under each degree configuration, we set the number of equally spaced intervals to $10$ and $20$, producing basis functions at a coarser scale and a finer scale, respectively. This resulted in a total of $70$ basis functions in $\bPhi$. In the higher-dimensional setting, we also used B-spline basis functions of degrees two and three but increased the number of equally spaced intervals to $5$, $15$, $25$, $\ldots$, $65$, 
yielding B-splines at seven different resolution scales. This higher-dimensional setting resulted in a total of $525$ basis functions in $\bPhi$. During implementation, we excluded basis functions that produced near-zero values in the columns of the $\bC$ matrix, which further reduced the number of effective basis functions from $525$ to $379$. Further details on the construction of B-spline basis functions can be found in Section 2 of Eilers \& Marx \cite{EilersandMarx1996}.

Based on the basis set $\bPhi$, we determined the neighborhood structure $\bG$ of basis functions in $\bPhi$ by examining the percentage of overlap in their supports, as described in Section \ref{sec:param}. We determined the initial value of $\bgamma$ in three steps: first, we obtained an initial estimate ${\beta}^0(t)$ using FLM-PACE; next, we performed Lasso regression by regressing ${\beta}^0(t)$ onto the elements of $\bPhi$, as described in Section \ref{sec:param}; and finally, we applied a hard threshold to the resulting regression coefficients to compute the initial value of $\bgamma$. For the Ising hyper-prior, we used the simplified two-parameter formulation and set the hyperparameters to $a = -7$, $d = 0.1$, and $v=44$. We applied Algorithm \ref{ag:flm} with $100,000$ MCMC iterations and used the first $50,000$ iterations as the burn-in period, which yielded $50,000$ posterior samples for $\bgamma, \sigma^2$ and $\bb$. Regarding the running time, it took approximately 6 minutes to complete $100,000$ iterations in the lower-dimensional setting and $1.8$ hours in the higher-dimensional setting. The computation was performed on a machine with an Intel64 Family 6 Model 126 processor @ 1.3 GHz CPU, 16 GB RAM, running Windows 11 Pro (64-bit). The convergence of the MCMC samples was assessed using trace plots and Geweke’s Z-statistics \cite{Geweke1992}. 
 
\begin{figure}
\centering
\includegraphics[width=5.5 in]{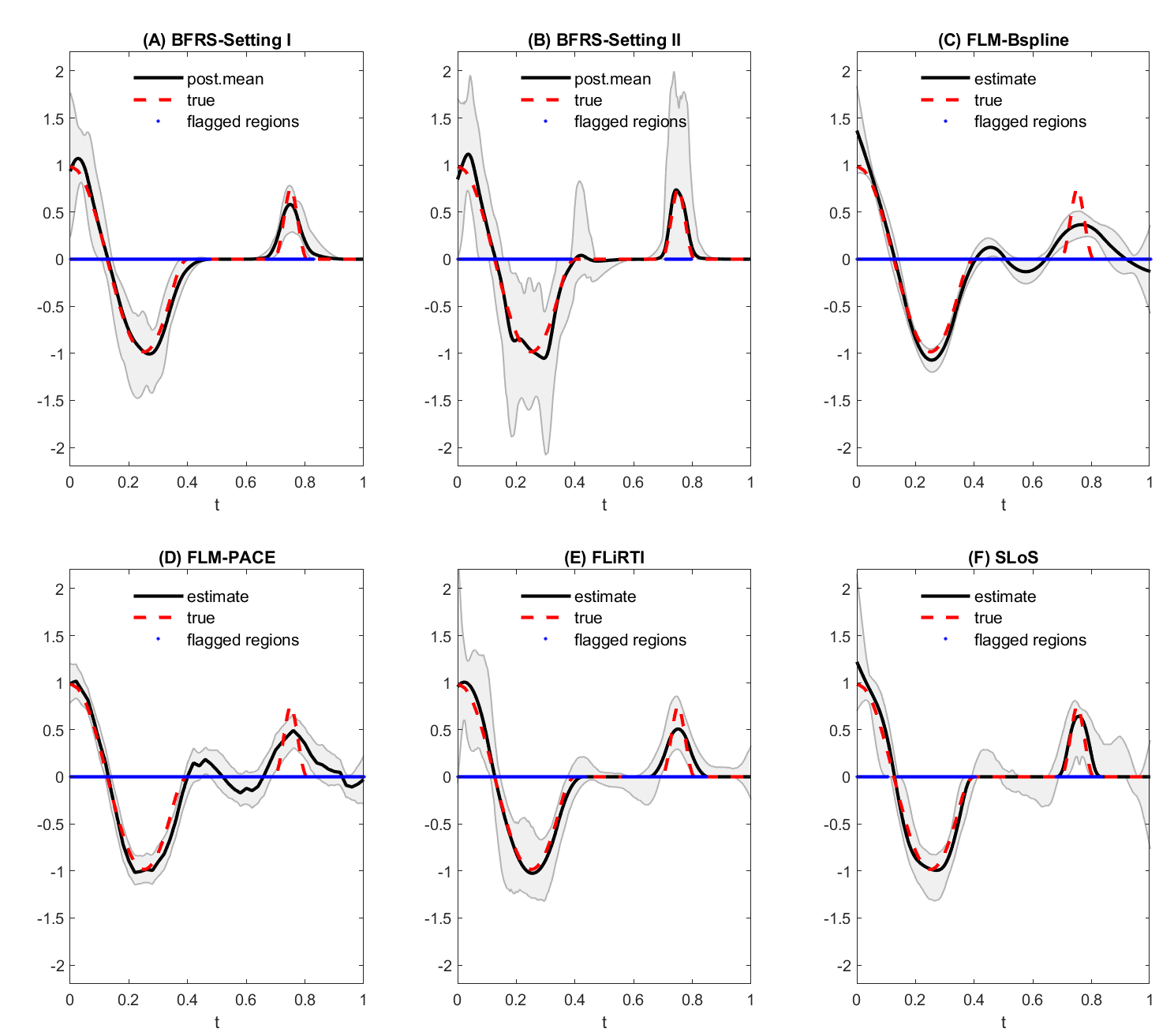}
\caption{Simulation results: Estimates of $\beta(t)$ using the proposed BFRS model under two different settings (A and B), compared with four other methods (C to F). Shaded areas in (A) and (B) represent the 95\% pointwise credible bands, while shaded areas in (C) to (F) represent the 95\% pointwise bootstrap confidence bands. In all plots, the red dashed lines represent the true $\beta(t)$, and the blue markings on the x-axis indicate the non-zero regions identified by the BFDR approach.} 
\label{simu_cont_response}
\end{figure}

Using posterior samples of $\bb$, we reconstructed posterior samples of $\beta(t)$ through the basis expansion $\beta^{(h)}(t) = \bPhi \bb^{(h)}$ for $h=1,\dots, H$, where $h$ is the index of the posterior samples. We calculated the posterior mean of $\beta(t)$ by averaging $\{\beta^{(h)}(t)\}$ pointwisely on the grid of $T$, and calculated the 95\% pointwise credible band by taking the $0.025$ and $0.975$ quantiles of the posterior samples pointwisely on the grid of $T$. Figure \ref{simu_cont_response} displays the posterior means of $\beta(t)$ (black lines), the 95\% pointwise credible bands (shaded areas), and the true $\beta(t)$ (red dashed lines) for the two basis settings in panels (A) and (B), respectively. In addition to estimating $\beta(t)$, we also applied the BFDR testing approach to identify its non-zero regions and highlighted the detected regions in blue on the x-axis in Figure \ref{simu_cont_response} (A) and (B). 

For comparison, we applied four alternative models mentioned previously: FLM-Bspline, FLM-PACE, FLiRTI, and SLoS. To apply FLM-Bspline, we followed the \texttt{fda} package by Ramsay et al. \cite{fdapackage}. The number of basis functions and the order of B-splines were selected from a list of candidates by minimizing the leave-one-out cross-validation score. 
For this simulation, we selected eight basis functions of order three (quadratic polynomial splines). For FLM-PACE, we used the \texttt{FPCreg} function of the \texttt{PACE} package \cite{PACE}. The number of functional principal components was determined to be eight, obtained by minimizing the regression Akaike Information Criterion (AIC). For the FLiRTI model, we applied the code shared by James et al. \cite{james2009functional}. The tuning parameters, including $\lambda$ and the weight placed on the zeroth derivative, were chosen based on cross-validation. The sparse solution was searched in the zeroth and the third derivatives which corresponds to piecewise quadratic polynomials. Finally, for SLoS, we applied the \texttt{SLOS} package shared by Lin et al. \cite{lin2015smooth}. We set the order of the B-spline basis to four, which produces cubic splines. We selected tuning parameters, including $\gamma$, $\lambda$ and the number of basis functions using AIC. For all comparison methods, once the tuning parameters were selected, we kept them fixed during the Bootstrap procedure. This decision was based on our observation that allowing the tuning parameters to vary across Bootstrap iterations resulted in much more unstable estimations and excessively wide confidence bands. 
In Figure \ref{simu_cont_response} (C) to (F), we plotted the estimated $\beta(t)$ (black lines) for the four comparison methods along with the 95\% pointwise Bootstrap confidence bands (shaded area) and the true $\beta(t)$ (red dashed lines). The Bootstrap samples of $\beta(t)$ also allowed us to detect non-zero regions using the BFDR testing approach. We highlighted the detected non-zero regions on x-axis with blue marks in Figure \ref{simu_cont_response} (C) to (F).

From Figure \ref{simu_cont_response} (A) and (B), we observe that both settings of the proposed BFRS model produced posterior means (black lines) that closely track the true $\beta(t)$ (red dashed lines), and both settings correctly identified the two non-zero regions (marked by blue) of $\beta(t)$. Among the two settings, setting II, which used $379$ basis functions compared to $72$ in setting I, resulted in a posterior mean that better tracks the higher-frequency region $[0.719, 0.781]$ of $\beta(t)$ than setting I. Furthermore, setting II also produced a wider credible band compared to setting I. Among the four comparison models, we observe that the two FLMs, FLM-Bspline and FLM-PACE, both produced $\beta(t)$ estimates (black lines) that are non-sparse. As a result, the BFDR testing approach identified all regions as non-zero. In contrast, the two sparse-encouraging methods, FLiRT and SLoS, produced sparse estimates of $\beta$ and correctly identified the non-zero regions. Compared with FLiRT, SloS appears to estimate the higher frequency region $[0.719, 0.781]$ of $\beta(t)$ better. Finally, both FLiRT and SLoS produced wider Bootstrap confidence bands for the zero regions and near the boundaries.   
   
In addition to Figure \ref{simu_cont_response}, we summarized the performance of estimation and region detection for $\beta(t)$ by calculating several summary statistics. The results are presented in the top pane of Table \ref{tab:simu_case1}. Specifically, four summary statistics related to the estimation of $\beta$ were considered: the Integrated Squared Error (ISE), calculated as $\mbox{ISE} = \int_T \big\{\widehat{\beta}(t) - \beta(t)\big\}^2dt$, where $\widehat{\beta}(t)$ denotes the posterior mean (or the $\beta(t)$ estimate if using a Frequentist method); the Integrated Posterior Variability (IPVar),  calculated as $\mbox{IPVar} = H^{-1}\sum_{h=1}^H\int_T \big\{{\beta}^{(h)}(t) - \widehat{\beta}(t)\big\}^2dt$; the Integrated Width (IWidth) of the 95\% pointwise credible band (or bootstrap confidence band), calculated as $\mbox{IWidth} = \int_T \left\{U(t) - L(t)\right\}dt$, where $U(t)$ and $L(t)$ are the upper and lower bounds of the $95\%$ credible band; and the Coverage Probability (CPr) for the $95\%$ Pointwise credible band, obtained by counting the proportion of $T$ on which the true $\beta(t)$ lies in the $95\%$ pointwise credible band. Among these summary statistics, ISE quantifies the discrepancy between the estimated $\beta(t)$ and the true value,
while IPVar characterizes the variability of the posterior (or bootstrap) samples around the posterior mean (or bootstrap mean). To evaluate the estimation of $\sigma$, we calculated two summary statistics: the Squared Error (SE) and the Posterior Variability (PVar). The SE is
defined as $(\widehat{\sigma}-\sigma)^2$ where $\widehat{\sigma}$ is calculated as the square root of the posterior mean (or other estimates if using non-Bayesian methods) of $\sigma^2$. The PVar is defined as $H^{-1}\sum_{h=1}^H ({\sigma}^{(h)} - \widehat{\sigma})^2$, where $\sigma^{(h)}$ represents the $h$th posterior sample of $\sigma$. SE measures the deviation of the estimate of $\sigma$ from the true value, while PVar captures the variability of the posterior (or bootstrap) samples around the posterior (or bootstrap) mean. To evaluate the performance of region detection using BFDR-based testing, we computed four additional summary statistics: (i) the false discovery rate (FDR), which reveals that among the discovery points (grid points of $T$ that are flagged as satisfying $|\beta(t)|>\delta$), which proportion is indeed false (i.e., grid points that truly satisfy $|\beta(t)|\leq \delta$); (ii) the false negative rate (FNR), which measures that, among those non-discovery points, which proportion truly satisfies $|\beta(t)|>\delta$; (iii) the sensitivity (SEN), obtained by calculating the proportion of correctly detected grid points among all grid points that truly satisfy $|\beta(t)|>\delta$; and (iv) the specificity (SPEC), which evaluates that, among the grid points that genuinely satisfy $|\beta(t)| \leq \delta$, which proportion is correctly identified as non-discovery points (not flagged). 

\begin{table}[h]
\caption {Simulation results: Comparison between two settings of BFRS (BFRS-I, BFRS-II) and several existing methods, including functional data regressions using Bspline or PACE, FLiRTI, and SLoS. The top section presents results for the continuous response case, while the bottom section presents results for the binary response case.  Note that for the binary response case, statistics for the $\sigma$ parameter are omitted as the model does not include the $\sigma$ parameter.
}
\vspace{0.1 in}
\label{tab:simu_case1}
\centering
\scalebox{0.83}{
\begin{tabular}{llcccccccccc}
\toprule
\toprule
Response& Method & \multicolumn{4}{c}{$\beta(t)$} & \multicolumn{2}{c}{$\sigma$ ($\times 1e-6$)} & \multicolumn{4}{c}{Region Detection Using BFDR} \\
\cmidrule(lr){3-6} \cmidrule(lr){7-8} \cmidrule(lr){9-12}
&  & ISE & IPVar & IWidth  & CPr & SE& PVar& FDR & FNR & SEN & SPEC\\
\midrule
& BFRS-I    & 0.005 & 0.021 & 0.355 & 0.915 &0.604&4.38& 0.110 & 0.015 & 0.983 & 0.900\\
& BFRS-II    & 0.006 & 0.059 & 0.696 & 1.00 &0.757&4.25& 0 & 0.058 & 0.941 & 1.00\\
& FLM-Bspline & 0.018 & 0.006 & 0.264 & 0.665 &4.58&3.55& 0.489 & 0 & 1.00 & 0.005\\
Continuous& FLM-PACE   & 0.017 & 0.007 & 0.334 & 0.810 &5.08& 3.95& 0.478 & 0 & 1.00 & 0.046\\
& FLiRTI    & 0.009 & 0.022 &  0.479 & 0.960 &9.56&3.88& 0.294 & 0 & 1.00 & 0.566\\
& SLoS     &  0.009 & 0.018 & 0.432 & 0.995 &0.583&4.61& 0.099 & 0.064 & 0.941 & 0.893\\\toprule
& BFRS   &  0.052 & 0.016 &  0.220 & 0.705 & --& -- & 0.124 & 0.220  & 0.760 & 0.888 \\
Binary & GFLM-Bspline   & 0.485 & 0.113 & 1.19 & 0.493 &--&--& 0.487 & 0.250 & 0.995 & 0.015\\
& GFLM-PACE  & 0.330 &0.060 & 0.913 & 0.363 &--&--& 0.486 & 0 & 1.00 &0.015\\
\bottomrule
\end{tabular}
}
\end{table} 

The results for all summary statistics are presented in the top portion of Table \ref{tab:simu_case1}. From Table \ref{tab:simu_case1}, we observe that, for the estimation of $\beta(t)$, both settings of the proposed BFRS model achieved smaller ISE values ($0.005$ and $0.006$) compared to all other methods. The two FLM models that do not assume sparsity in $\beta(t)$, FLM-Bspline and FLM-PACE, resulted in larger ISEs ($0.018$ and $0.017$, respectively) than the models that assume sparsity. Furthermore, FLM-Bspline and FLM-PACE also produced smaller IPVar ($0.006$ and $0.007$), narrower IWidth ($0.264$ and $0.334$), and lower CPr values ($0.665$ and $0.810$) for $\beta(t)$. These results suggest that, when the true $\beta(t)$ contains zero regions, incorporating a sparse assumption for $\beta(t)$ leads to reduced estimation error and improved coverage probability of the confidence band. Among the models that assume sparse $\beta(t)$, the two settings of the proposed BFRS model achieved smaller ISE for $\beta(t)$ ($0.005$ and $0.006$) than FLiRTI and SLoS ($0.009$). Overall, BFRS-I exhibited comparable IPVar with FLiRTI and SLoS ($0.021$ vs. $0.022$ and $0.018$), but yielded narrower IWidth ($0.355$ vs. $0.479$ and $0.432$) and lower CPr ($0.915$ vs. $0.960$ and $0.995$). In contrast, BFRS-II, which employs a larger number of basis functions than BFRS-I ($379$ vs. $72$), resulted in the highest IPVar, Iwidth and CPr among all methods. This pattern suggests that using excessive amounts of redundant basis functions may lead to increased posterior variance, wider confidence band, and higher coverage probability. For the estimation of $\sigma$, while all methods yielded very small SE and PVar on the order of $1e-6$,  three of the methods---FLM-Bspline, FLM-PACE and FLiRTI---produced relatively larger SE than others.

In addition to the estimation results, Table \ref{tab:simu_case1} also presents statistics on region detection using the BFDR testing approach. From these statistics, we observe that the two methods that do not assume sparsity in $\beta(t)$---FLM-Bspline and FLM-PACE---resulted in very high sensitivity ($1.00$ for both cases) but very low specificity ($0.005$ and $0.046$, respectively). Consequently, their true false discovery rates were higher than those of other methods ($0.489$ and $0.478$). This outcome is expected, as both methods produce $\beta(t)$ estimates that are non-zero everywhere. As a result, the BFDR approach identified all regions of $\beta(t)$ as non-zero. For the two Frequentist methods that assume sparsity in $\beta(t)$ (FLiRTI and SLos), while they yielded improved specificity ($0.566$ and $0.893$) compared to FLM-Bspline and FLM-PACE, all statistics remained inferior to those of the two settings of the proposed BFRS. Finally, comparing the two settings of the proposed model, BFRS-I and BFRS-II, we observe that BFRS-II achieved a lower false discovery rate (0 vs. $0.110$), a higher false negative rate ($0.058$ vs. $0.015$), lower sensitivity ($0.941$ vs $0.983$), and higher specificity ($1.00$ vs. $0.900$). These statistics suggest that, with the same prior parameter setup, using a larger number of basis functions may result in sparser estimates of $\beta$, identifying more zero regions and fewer non-zero regions. 
Overall, both settings of the proposed BFRS model produced region detection results that outperformed those of the other methods. Among the Frequentist methods, SLoS exhibited performance closest to that of our model. 

\subsection{The binary response case}
For the binary response case, we applied the proposed BFRS model using a set of over-complete B-spline bases at two degrees: degree two and three, corresponding to quadratic and cubic polynomials, respectively. For each degree, we adopted two different resolution scales, similar to the setup in setting I for the continuous response case. This resulted in a total of $70$ basis functions. Initial values for $\bb$ and $\bgamma$ were obtained by applying the continuous response model with random initialization. The prior hyperparameters were set to $a = -4$, $d = 0.1$ and $v = 40$. A total of $200,000$ MCMC iterations were conducted, with the first $100,000$ iterations designated as the burn-in period. Regarding the running time, it took approximately $48$ minutes to complete $200,000$ MCMC iterations under the current setting. The convergence of the MCMC samples was assessed using trace plots and Geweke’s Z-test, as in the continuous response case. Based on the posterior samples, we calculated summary statistics for $\beta(t)$ and the region selection statistics following the same approach as in the continuous response case. For comparison, we applied two GFLM methods: one utilizing B-splines, denoted as GFLM-Bspline \cite{Febrero2012}, and the other based on functional principal component analysis, denoted as GFLM-PACE \cite{mueller2005}. In both methods, the logit link function was used. Summary statistics for these two GFLM methods were calculated based on $1,000$ bootstrap estimates of $\beta(t)$. The results were reported in the bottom panel of Table \ref{tab:simu_case1}.

The results of the proposed BFRS, shown in the bottom panel of Table \ref{tab:simu_case1}, reveal that 
estimating $\beta(t)$ is more challenging in the binary response case compared to the continuous response case. This is evident from the larger ISE ($0.052$ vs. $0.005$ and $0.006$), lower coverage probability ($0.705$ vs. $0.915$ and $1.00$), as well as reduced sensitivity ($0.760$ vs. $0.983$ and $0.941$) and specificity ($0.888$ vs. $0.900$ and $1.00$) in region selection. This outcome is not surprising because, under the same sample size, binary responses contain substantially less information about $\beta(t)$ than continuous responses, making the estimation of $\beta(t)$ more difficult. The two GFLM methods, GFLM-Bspline and GFLM-PACE, do not assume zero regions in $\beta(t)$, and thus lead to non-zero $\beta(t)$ estimates across nearly the entire domain of $T$. As a result, the BFDR region detection method yielded very high sensitivity ($0.995$ and $1.00$) and very low specificity ($0.015$), indicating that nearly all regions of $\beta(t)$ were identified as non-zero. Additionally, these methods also resulted in higher ISE ($0.485$ and $0.330$ vs. $0.052$) and lower coverage probability ($0.493$ and $0.363$ vs. $0.705$) for the estimation of $\beta(t)$ compared to the proposed BFRS.

\section{Real data application}
\label{sec:realdata}

\begin{figure} 
\centering
\includegraphics[width=5.4 in]{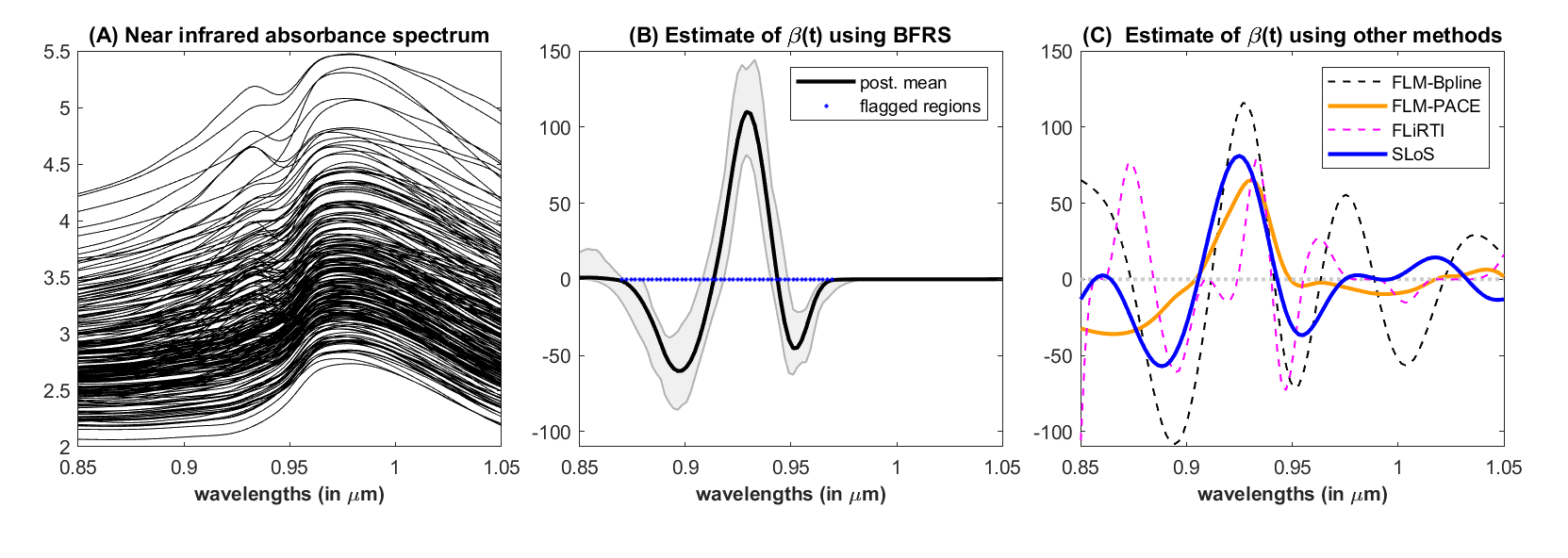}
\vspace{-0.2 in}
\caption{\label{fig:realdataplot} Real data application: (A) Plot of the $240$ near-infrared spectral curves; (B) Posterior mean of $\beta(t)$ and the 95\% pointwise credible band, with regions where $\beta(t)>1e-3$ flagged by blue marks on the x-axis; (C) Estimates of $\beta(t)$ obtained using four other methods.}
\end{figure}

We first demonstrate the effectiveness of the proposed BFRS method by applying it to the Tecator dataset. This dataset contains near-infrared absorbance spectra of $240$ meat samples measured at 100 channels (i.e., 100 grid points), with wavelengths ranging from 850 nanometer (nm) to 1,050 nm. Along with the spectra, the percentages of moisture (water), fat, and protein in each meat sample, determined by analytic chemistry, are also provided. The $240$ samples are pre-divided into five subsets, including $129$ training samples, $43$ monitoring samples, $43$ testing samples, and $25$ extrapolation samples. The dataset was collected by the Tecator company and is publicly available on the StatLib website (\url{https://lib.stat.cmu.edu/datasets/tecator}). A plot of the $240$ spectral curves is illustrated in Figure \ref{fig:realdataplot} (A). The original goal of the data providers was to use the spectral measurements to predict the fat content. In our analysis, in addition to prediction, we also hope to find out which regions of the near-infrared spectrum are most relevant to the fat content. Based on suggestions from the data providers, we designated the first $172$ samples as the training set, comprising the first two pre-defined subsets: the training samples and the monitoring samples. Additionally, we utilized the $43$ testing samples as the test set for out-of-sample predictions. We treated the spectral curves as the functional predictors, denoted by $\{X_i(t)\}$, where $t$ is scaled in micrometers ($\mu m$), i.e. $t\in [0.85 ~\mu m, 1.05 ~\mu m]$. The percentages of fat were scaled to the range $[0,1]$ and treated as the scalar responses.

We applied the proposed BFRS method using Algorithm \ref{ag:flm}. In the implementation, we adopted two sets of B-spline bases (L=2) with $(q_1 = 2, k_1 = 10)$ and $(q_2 = 3, k_2 = 20)$. This resulted in a total of $70$ basis functions at two different scales and two different degrees. The hyperparameters were set to $a = -6$, $d = 0.1$, and $v = 1e+3$, following suggested procedures in Section \ref{subsec:ising}. The initial value of $\bgamma$ was obtained by sampling from independent Bernoulli distributions with a success probability $0.5$. A total of $20,000$ MCMC iterations were conducted, with the first $10,000$ iterations designated as the burn-in period. It took approximately $50$ seconds to run $20,000$ MCMC iterations. Convergence diagnostics were evaluated using trace plots and Geweke’s Z-test. Based on the posterior samples, the posterior mean of $\beta(t)$ and the 95\% pointwise credible band were computed. Regions where $\beta(t)>\delta$ were identified using the BFDR approach, with $\delta=1e-3$. Results are presented in Figure \ref{fig:realdataplot} (B).

From Figure \ref{fig:realdataplot} (B), we observe two zero regions on the estimate of $\beta(t)$, the initial region $[0.850, 0.868]~\mu m$ and the ending region $[0.971, 1.050]~\mu m$. This indicates that spectral measurements in these two regions are not strongly associated with the fat content of the meat samples. Additionally, the estimate of $\beta(t)$ is positive in the region $[0.918, 0.940]~\mu m$ and negative in two regions, $[0.880,0.906]~\mu m$ and $[0.949,0.967]~\mu m$. This suggests that higher near-infrared absorbance intensity in the interval $[0.918, 0.940]~\mu m$ is associated with higher fat content, while in the intervals $[0.880,0.906]~\mu m$ and $[0.949,0.967]~\mu m$, it is associated with lower fat content. 

Similar to the simulation studies, we compared the performance of BFRS with four existing methods: FLM-Bspline, FLM-PACE, FLiRTI, and SLoS. When implementing these methods, most tuning parameters were selected using five-fold cross-validation on the training set. These tuning parameters included the number of basis functions and the roughness penalty for FLM-Bspline and SLoS, the weight and $\lambda$ parameter for FLiRTI, and the sparseness penalty for SLoS. For FLM-PACE, the number of functional principal components was chosen to ensure that the fraction of variance explained (FVE) was at least $99.9\%$. The estimated $\beta(t)$ was shown in Figure \ref{fig:realdataplot} (C). From Figure \ref{fig:realdataplot} (C), we observe that, similar to the BFRS estimate, all four methods produced $\beta(t)$ estimates with a peak around $0.930~\mu m$. Additionally, three methods---FLM-Bspline, FliRTI, and SloS---exhibited two valleys around $0.890~\mu m$ and $0.950~\mu m$. Furthermore, FLiRTI and SLoS generated close-to-zero estimates around $0.86~\mu m$, while FLM-PACE, FLiRTI and SLoS produced close-to-zero estimates in the region $[0.97,1.05]~\mu m$. Despite these similarities, none of the four comparison methods produced exact zero regions in their $\beta(t)$ estimates. Overall, we found that the estimate from SLoS most closely resembles the result of BFRS.

Using the estimates obtained from the training set, we predicted the fat content of test samples and 
reported the Mean Squared Error (MSE) in Table \ref{tab:tecator_mse}. 
\begin{table}[h]
\caption {Real data application: Prediction results on the testing samples.}
\vspace{0.1 in}
\label{tab:tecator_mse}
\centering
\begin{tabular}{lccccc}
\toprule
\toprule
Methods & BFRS & FLM-Bspline & FLM-PACE & FLiRTI & SLoS\\
\midrule
MSE ($\times1e-3$)    & 1.10 & 0.81 & 2.10 & 0.69 &1.40\\
\bottomrule
\end{tabular}
\end{table} 
Table \ref{tab:tecator_mse} shows that the proposed BFRS method yielded MSE higher than those of FLM-Bspline and FLiRTI ($1.10e-3$ vs. $0.81e-3$ and $0.69e-3$), but lower than those of FLM-PACE and SLoS ($2.10e-3$ and $1.40e-3$). These results suggest that, for the Tecator data, while encouraging sparsity in regions of $\beta(t)$ provides interpretative advantages, it does not lead to significant improvements in prediction accuracy compared to the other methods.

\section{Discussion}
\label{sec:discuss}

We proposed a Bayesian method for detecting regions on functional data that are relevant to a variable of interest. The proposed method achieves region selection through the sparse estimation of the regression coefficient function in a functional regression framework. This framework accommodates both continuous and binary response cases. Unlike existing methods that rely on Frequentist approaches and provide only point estimates, our approach offers full Bayesian inference, enabling a natural characterization of uncertainty in model parameters. A key contribution of the proposed method lies in its flexibility to accommodate over-complete basis functions, enabling the adaptive selection of heterogeneous local regions at different scales and locations through multi-scale basis functions. To promote continuous sparse estimation of the regression coefficient function, we integrate a spike-and-slab prior in the basis space with an Ising hyper-prior, effectively accounting for the neighborhood structure among basis functions. This idea of region selection for functional data can be broadly extended to a wider class of latent variable models. For instance, the latent variable $Z_i$ in equation (\ref{eq:latent}) could be linked to other types of responses, such as categorical responses or count data.
  
In the proposed framework, we assumed that the functional predictors $\{X_i\}$ are fully observed. When $\{X_i\}$ are subject to measurement errors,  model (\ref{eq:flm}) can still be represented in the form of (\ref{eq:compact_model}). Specifically, suppose that $W_i(t)$ is observed such that $W_i(t)=\widetilde{X}_i(t)+\delta_i(t)$, where $\delta_i(t)$ denotes the measurement error. We can represent $\widetilde{X}_i$ using a set of basis $\bPsi=(\varphi_1,\dots, \varphi_m)$ so that $\widetilde{X}_i=\bPsi\boldsymbol{x}_i$, where $\boldsymbol{x}_i$ is a vector of basis coefficients. Consequently, we can formulate the $\bC$ matrix in (\ref{eq:compact_model}) as $\bC=\left(\boldsymbol{1}, \boldsymbol{X}\bXi\right)$, where $\boldsymbol{X}$ is a matrix with rows formed by stacking $\boldsymbol{x}_i$ and $\bXi=(\langle \varphi_i, \phi_j^l \rangle)$. The error term $\epsilon_i$ in (\ref{eq:compact_model}) is then replaced by $\tilde{\epsilon}_i=\int_T \delta_i(t)\beta(t) dt +\epsilon_i$. With this modified $\bC$, the subsequent inference steps and algorithms remain unchanged. 

In our simulation and real data application, we used a simplified parameterization of the Ising hyper-prior, with the parameters $(a, d)$ determined following the recommendations of Li and Zhang \cite{li2010bayesian}. Alternative approaches for determining the hyperparameters have been discussed by Li and Zhang \cite{McEvoy2013}. In addition to the approaches discussed in Section \ref{subsec:ising}, commonly used techniques such as cross-validation can also be applied to determine these parameters.

In this paper, we focused on functional data defined on one-dimensional domains, i.e. $T \in \mathbb{R}$. The proposed framework is also applicable to functional data on higher-dimensional domains, such as $\mathbb{R}^2$ or $\mathbb{R}^3$. The extension can be accomplished by utilizing compactly supported bases, such as tensor product B-splines, wavelets, or finite elements, in $2D$ or $3D$. Additionally, while various techniques have been implemented to enhance the computational efficiency of the proposed algorithms, further improvements could be made by leveraging other fast stochastic search methods, such as the shotgun algorithm \cite{hans2007}. Finally, to enhance the mixing of the MCMC algorithms, one can consider adopting the truncated sparsity prior \cite{yang2016} as an alternative to the spike-and-slab prior.

The MATLAB code for the proposed algorithms has been made available to the readers in the online supplementary materials.



%

%

\bibliography{region_selection}

@article{Geweke1992,
   author = {J. Geweke},
   title = {Evaluating the accuracy of
sampling-based approaches to the calculation of
posterior moments},
   journal = {Bayesian Statistics 4},      
   year = {1992}   
}

@article{Olshausen1996,
  author = {Olshausen, Bruno A. and Field, David J.},
  title = {Emergence of simple-cell receptive field properties by learning a sparse code for natural images},
  journal = {Nature},
  year = {1996},
  volume = {381},
  pages = {607-609}
}

@article{zeng2024bayesian,
  title={Bayesian Image-on-Scalar Regression with a Spatial Global-Local Spike-and-Slab Prior},
  author={Zeng, Zijian and Li, Meng and Vannucci, Marina},
  journal={Bayesian Analysis},
  volume={19},
  number={1},
  pages={235--260},
  year={2024},
  publisher={International Society for Bayesian Analysis}
}

@article{ZHU2018501,
author = {Hongxiao Zhu and Francesco Versace and Paul M. Cinciripini and Philip Rausch and Jeffrey S. Morris},
title = {Robust and Gaussian spatial functional regression models for analysis of event-related potentials},
journal = {NeuroImage},
volume = {181},
pages = {501-512},
year = {2018},
issn = {1053-8119}
}

@article{Huo2022,
author = {Shuning Huo and Jeffrey S Morris and Hongxiao Zhu},
title = {Ultra-Fast Approximate Inference Using Variational Functional Mixed Models},
journal = {Journal of Computational and Graphical Statistics},
volume = {32},
number = {2},
pages = {353--365},
year  = {2023},
publisher = {Taylor & Francis}
}

@misc{PACE,
  author = {Fang Yao and Bitao Liu and Wenwen Tao and Shuang Wu and Nuen Tsang Yang and Wenjing Yang and Hans-Georg Müller and Jane-Ling Wang},
  title = {{PACE:} Principal Analysis by Conditional Expectation},
  howpublished = {\url{https://hadoop.apache.org}},
  note = {Version 2.17},
	year = {2015}  
}

@Manual{fdapackage,
    title = {fda: Functional Data Analysis},
    author = {J. O. Ramsay and Spencer Graves and Giles Hooker},
    year = {2022},
    note = {R package version 6.0.3},
    url = {https://cran.r-project.org/package=fda}		
  }

@article{EilersandMarx1996,
author = {Paul H. C. Eilers and Brian D. Marx},
title = {{Flexible smoothing with B-splines and penalties}},
volume = {11},
journal = {Statistical Science},
number = {2},
publisher = {Institute of Mathematical Statistics},
pages = {89 -- 121},
year = {1996}
}

@article{Handcock94,
author = { Mark S.   Handcock  and  James R.   Wallis },
title = {An Approach to Statistical Spatial-Temporal Modeling of Meteorological Fields},
journal = {Journal of the American Statistical Association},
volume = {89},
number = {426},
pages = {368-378},
year  = {1994},
publisher = {Taylor & Francis}
}

@article{LuLou2022,
author = {Zihang Lu and Wendy Lou},
title = {Bayesian approaches to variable selection: a comparative study from practical perspectives},
journal = {The International Journal of Biostatistics},
number = {1},
volume = {18},
year = {2022},
pages = {83--108}
}

@article{OHara2009,
author = {R. B. O'Hara and M. J. Sillanpää},
title = {{A review of Bayesian variable selection methods: what, how and which}},
volume = {4},
journal = {Bayesian Analysis},
number = {1},
publisher = {International Society for Bayesian Analysis},
pages = {85 -- 117},
year = {2009}
}

@phdthesis{Weller2014,
    author     =     {Weller, Adrian},
    title     =     {Methods for Inference in Graphical Models},
    school     =     {Columbia University},
    address     =     {New York},
    year     =     {2014}
    }

@book{ramsay2009introduction,
  title={Introduction to functional data analysis},
  author={Ramsay, JO and Hooker, Giles and Graves, Spencer},
  booktitle={Functional data analysis with R and MATLAB},
  pages={1--19},
  year={2009},
  publisher={Springer}
}

@article{Wolfinger1994,
	author={Wolfinger, R. and Tobias, R. and Sall, J.},
	title={Computing Gaussian Likelihoods and Their Derivatives for General Linear Mixed Models},
	journal={SIAM Journal on Scientific Computing},
	year={1994},
	volume={15},
	pages={1294-1310}
}

@inproceedings{HeNIPS2016,
 author = {He, Bryan D and De Sa, Christopher M and Mitliagkas, Ioannis and R\'{e}, Christopher},
 booktitle = {Advances in Neural Information Processing Systems},
 editor = {D. Lee and M. Sugiyama and U. Luxburg and I. Guyon and R. Garnett},
 pages = {},
 publisher = {Curran Associates, Inc.},
 title = {Scan Order in Gibbs Sampling: Models in Which it Matters and Bounds on How Much},
 volume = {29},
 year = {2016}
}

@book{Ramsay1997,
	author={Ramsay, J. O. and Silverman, B. W.},
	title={Functional Data Analysis},
	publisher={Springer-Verlag},
	year={1997},
	address={New York}
}

@ARTICLE{Wang2016,
   author = {Wang, J.-L. and Chiou, J.-M. and M\"uller, H. G.},
    title = "Functional data analysis",
  journal = {The Annual Review of Statistics and Its Application},
     year = 2016,
     volume=3,
     number = 1,
     pages={257-295}
}

@article{McEvoy2013,
author = "McEvoy, B. W. and Nandy, R. R. and Tiwari, R. I.",
journal = "Biometrics",
pages = "661--672",
title = "Bayesian Approach for Clinical Trial Safety Data Using an {I}sing Prior",
volume = "69",
year = "2013"
}

@article{yang2016,
author = "Yang, Yun and Wainwright, Martin J. and Jordan, Michael I.",
journal = "The Annals of Statistics",
number = "6",
pages = "2497--2532",
publisher = "The Institute of Mathematical Statistics",
title = "On the computational complexity of high-dimensional Bayesian variable selection",
volume = "44",
year = "2016"
}

@article{hans2007,
author = {Chris Hans and Adrian Dobra and Mike West},
title = {Shotgun Stochastic Search for Large p Regression},
journal = {Journal of the American Statistical Association},
volume = {102},
number = {478},
pages = {507-516},
year  = {2007},
publisher = {Taylor & Francis}
}

@article{Morris2008,
	author={Morris, J. S. and Brown, P. J. and Herrick, R. C. and Baggerly,  K. A. and Coombes, K. R.},
	title={Bayesian analysis of mass spectrometry proteomic data using wavelet-based functional mixed models},
	journal={Biometrics},
	year={2008},
	volume={64},
	pages={479-489}
}

@article{Meyer2015,
	author={Meyer, M. J. and Coull, B. A. and Versace, F. and Cinciripini, P. and Morris, J. S.},
	title={Bayesian function-on-function regression for multilevel functional data},
	journal={Biometrics},
	date={2015 Mar 18},
	volume={71},
	number={3},
	pages={563-74},
	year={2015}
	}

@article{mueller2005,
author = "M{\"u}ller, Hans-Georg and Stadtm{\"u}ller, Ulrich",
fjournal = "The Annals of Statistics",
journal = "The Annals of Statistics",
number = "2",
pages = "774--805",
publisher = "The Institute of Mathematical Statistics",
title = "Generalized functional linear models",
volume = "33",
year = "2005"
}

@article{Febrero2012,
   author = {Manuel Febrero-Bande and Manuel de la Fuente},
   title = {Statistical Computing in Functional Data Analysis: The {R} Package fda.usc},
   journal = {Journal of Statistical Software},
   volume = {51},
   number = {4},
   year = {2012},
   issn = {1548-7660},
   pages = {1--28}
}

@incollection{Vannucci2012,
  author      = "Marina Vannucci and Francesco C. Stingo",
  title       = "Bayesian Models for Variable Selection that Incorporate Biological Information",
  editor      = "Bernardo, J. M. and Bayarri, M. J. and Berger, J. O. and Dawid, A. P. and Heckerman, D. and Smith, A. F. M. and West, M.",
  booktitle   = "Bayesian Statistics 9",
  publisher   = "Oxford Scholarship Online",
  address     = "Oxford",
  year        = 2012,
  pages       = "1--29"
}

@book{Ferraty2006,
    author = {Ferraty, Fr\'{e}d\'{e}ric and Vieu, Philippe},
    publisher = {Springer Verlag},
    title = {Nonparametric Functional Data Analysis: Theory and Practice},
    year = {2006}
}

@book{Horvath2012,
    author = {Horv\'{a}th, Lajos and Kokoszka, Piotr},
    publisher = {Springer Verlag},
    title = {Inference for Functional Data with Applications},
    year = {2012}
}

@article{li2010bayesian,
  title={Bayesian variable selection in structured high-dimensional covariate spaces with applications in genomics},
  author={Li, Fan and Zhang, Nancy R},
  journal={Journal of the American statistical association},
  volume={105},
  number={491},
  pages={1202--1214},
  year={2010},
  publisher={Taylor \& Francis}
}

@article{lin2015smooth,
  title={Locally Sparse Estimator for Functional Linear Regression Models},
  author={Lin, Zhenhua and Cao, Jiguo and Wang, Liangliang and Wang, Haonan},
  journal={Journal of Computational and Graphical Statistics},
	volume={26},
  number={2},
	pages={306–318},
  year={2017}
}

@article{koltchinskii2013l_1,
  title={{L1}-Penalization in Functional Linear Regression with Subgaussian Design},
  author={Koltchinskii, Vladimir and Minsker, Stanislav},
  journal={Journal de l{\textquoteright}\'Ecole polytechnique - Math\'ematiques},
  pages = {269--330},
  publisher = {Ecole polytechnique},
  volume = {1},
  year = {2014}
}

@article{zhou2013functional,
  title={Functional linear model with zero-value coefficient function at sub-regions},
  author={Zhou, Jianhui and Wang, Nae-Yuh and Wang, Naisyin},
  journal={Statistica Sinica},
  volume={23},
  number={1},
  pages={25},
  year={2013},
  publisher={NIH Public Access}
}

@article{james2009functional,
  title={Functional linear regression that's interpretable},
  author={James, Gareth M and Wang, Jing and Zhu, Ji},
  journal={The Annals of Statistics},
  pages={2083--2108},
  year={2009},
  publisher={JSTOR}
}

@article{Morris2015,
  title={Functional regression},
  author={Morris, Jeffrey S},
  journal={Annual Review of Statistics and Its Application},
  volume={2},
  pages={321--359},
  year={2015},
  publisher={Annual Reviews}
}

@book{ramsay2006functional,
  title={Functional data analysis},
  author={Ramsay, James O},
  year={2006},
  publisher={Wiley Online Library}
}

@article{yao2005functional,
  title={Functional data analysis for sparse longitudinal data},
  author={Yao, Fang and M{\"u}ller, Hans-Georg and Wang, Jane-Ling},
  journal={Journal of the American Statistical Association},
  volume={100},
  number={470},
  pages={577--590},
  year={2005},
  publisher={Taylor \& Francis}
}

\bmsection*{Supporting information}

Additional supporting information may be found in the
online version of the article at the publisher’s website.

%

\end{document}